\documentclass[pdflatex,sn-mathphys-num]{sn-jnl}% Math and Physical Sciences Numbered Reference Style
\usepackage{graphicx}%
\usepackage{multirow}%
\usepackage{amsmath,amssymb,amsfonts}%
\usepackage{amsthm}%
\usepackage{mathrsfs}%
\usepackage[title]{appendix}%
\usepackage{xcolor}%
\usepackage{textcomp}%
\usepackage{manyfoot}%
\usepackage{booktabs}%
\usepackage{algorithm}%
\usepackage{algorithmicx}%
\usepackage{algpseudocode}%
\usepackage{listings}%
\usepackage{amsmath, amsthm, amssymb, amsfonts}
\usepackage{graphicx}
\usepackage{caption}
\usepackage{subcaption}
\usepackage{tabularx}
\usepackage{xfrac}

\renewcommand{\orcidlogo}{%
  \includegraphics[width=10pt]{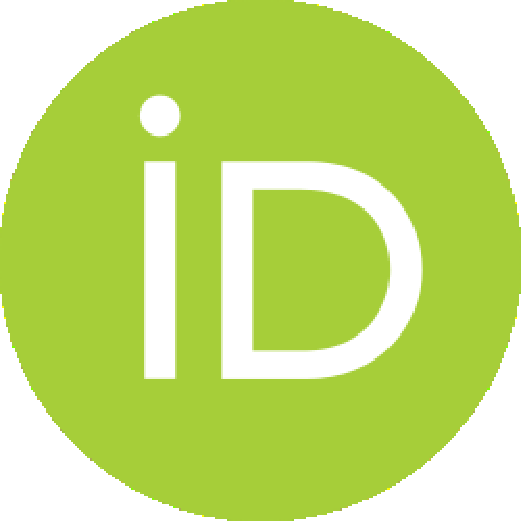}%
}

\renewcommand{\orcid}[1]{%
  \href{https://orcid.org/#1}{\orcidlogo}%
}
\usepackage{hyperref}
\hypersetup{colorlinks,
citecolor=blue, urlcolor=blue, linkcolor=blue}
\usepackage{upgreek}
\usepackage[greek,english]{babel}
\usepackage[utf8]{inputenc}

\usepackage[capitalise]{cleveref}
\usepackage{multirow}
\newcommand{\munu}{{\mu \nu}}

\theoremstyle{thmstyleone}%
\theoremstyle{thmstyletwo}%

\theoremstyle{thmstylethree}%

\begin{document}

\title[Critical analysis of some weak-field Weyl conformal gravity solutions]{Critical analysis of some weak-field Weyl conformal gravity solutions}


\author*[1,2]{\fnm{Miguel} \sur{Yulo Asuncion}\orcid{0009-0003-5075-3107}}\email{jesusmiguel.yuloasuncion@nottingham.ac.uk}

\author[3]{\fnm{Reinosuke} \sur{Kusano}\orcid{0000-0001-7569-7197}}\email{rk77@st-andrews.ac.uk}

\affil*[1]{\orgdiv{Nottingham Centre of Gravity}, \orgaddress{ \city{Nottingham}, \postcode{NG7 2RD},  \country{United Kingdom}}}

\affil[2]{\orgdiv{School of Mathematical Sciences}, \orgname{University of Nottingham}, \orgaddress{ \city{Nottingham}, \postcode{NG7 2RD}, \country{United Kingdom}}}

\affil[3]{\orgdiv{SUPA Physics and Astronomy}, \orgname{University of St\,Andrews}, \orgaddress{\city{St Andrews}, \postcode{KY16 9SS}, \state{Scotland}, \country{United Kingdom}}}


\abstract{The field equations of Conformal Gravity (CG) are constructed from fourth-order derivatives of the metric. While the second-order field equations of General Relativity (GR) reduce to a second-order Poisson equation in the weak-field limit, those of CG yield a fourth-order one. Despite this, works by Tanhayi, Fathi, and Takook~[{Mod. Phys. Lett. A} \textbf{26}, 2403 (2011)], Payandeh and Fathi~[{Int. J. Th. Phys.} \textbf{51}, 2227 (2012)], and Fathi, Olivares, and Villanueva~[{Galaxies} \textbf{9}, 43 (2021)] use second-order Poisson equations to generate purported solutions to CG in a weak-field limit. We demonstrate that these solutions are erroneous due to the use of second-order Poisson equations instead of appropriate fourth-order ones. We also point out other inconsistencies in the derivations of these metrics, such as the assertion that known CG (electro)vacuum solutions describe the interiors of matter distributions.}




\maketitle

\section{Introduction}

Weyl Conformal Gravity (CG)~\cite{Weyl1918, Bach1921} is a proposed alternative to Einstein's General Relativity (GR)~\cite{Einstein:1915ca} that has been used to account for flat galactic rotation curves~\cite{flat, flat2} and the accelerating expansion of the universe~\cite{riessexpansion,perlmutter1999} 
without recourse to the existence of dark matter and dark energy~\cite{mannheim1989, mannheim2011, Mannheim2012Rot, Mannheim2013Rot, KeithRotation,VarieschiParameters1, mannheimgammavalue, VarieschiCosmo, mannheim_2022}. It also generates some interest as a prospective quantum gravity theory due to its renormalizability~\cite{mannheim2009, makingthecase, grumiller2014}, and has been shown to be able to avoid Ostrogradsky ghosts despite being a higher-order theory~\cite{bender2008_1, bender2008_2, bender2008_3}. 

As its name suggests, in addition to the invariance to both the coordinate $g_{\mu \nu}(x) \rightarrow g'_{\mu \nu} (x')$ and Lorentz $x^\mu \rightarrow \Lambda^\mu_\nu \ x^\nu$ transformations found in GR, CG bears an additional conformal symmetry as well. This is an invariance to local conformal transformations  $g_{\mu \nu}(x) \rightarrow \Tilde{g}_{\mu \nu}(x) = \Omega^2(x) g_{\mu 
\nu}(x)$~\cite{turner2020}, where $\Omega(x)$ is a conformal stretching factor~\cite{VarieschiParameters1}. This conformal invariance is built into the Weyl action
\begin{equation}\label{eq:Weyl_action}
    I_{\rm W} = -\alpha_\mathrm{g} \int \text{d}^4 x \,\sqrt{-g}\, C_{\lambda \mu \nu \kappa}\,C^{\lambda \mu \nu \kappa}\ ,
\end{equation}
where $g$ is the determinant of the metric $g_{\mu\nu}$
and $\alpha_{\text{g}}$ is a dimensionless gravitational coupling constant. To recover attractive gravity in the Newtonian limit, this coupling is taken to be negative $\alpha_\mathrm{g} < 0$~\cite{mannheim_2007}. Meanwhile, $C_{\lambda \mu \nu \kappa}$ is the conformal Weyl tensor, which is the traceless part of the Riemann tensor $R_{\lambda \mu \nu \kappa}$ given as 
\begin{equation}\label{eq:Weyl_tensor}
\begin{aligned}
C_{\lambda \mu \nu \kappa} = R_{\lambda \mu \nu \kappa} -& \dfrac{1}{2}(g_{\lambda\nu} R_{\mu\kappa} - g_{\lambda\kappa}R_{\mu\nu} \notag - g_{\mu\nu}R_{\lambda\kappa}+g_{\mu\kappa}R_{\lambda\nu}) \notag \\&+ \dfrac{1}{6}R(g_{\lambda\nu}g_{\mu\kappa} - g_{\lambda\kappa}g_{\mu\nu}),
\end{aligned}
\end{equation}
where $R_{\mu \nu}$ is the Ricci tensor and $R$ the Ricci scalar. Like the Riemann tensor, the Weyl tensor contains second-order derivatives of the metric $g_{\mu \nu}$. Variation of this action~\eqref{eq:Weyl_action} with respect to $g_{\mu \nu}$ leads to the Bach field equations~\cite{Bach1921, mannheimgammavalue} 
\begin{equation}\label{eq:BFE}
    W_{\mu\nu} =\dfrac{1}{4\,\alpha_\mathrm{g}} T_{\mu\nu}.
\end{equation}
Here, $T_{\mu \nu}$ is the stress-energy tensor, and
\begin{equation}\label{eq:Bach Tensor}
W_{\mu \nu}=2\,\nabla^\kappa\nabla^\lambda C_{\mu\lambda\nu\kappa}
-
R^{\lambda\kappa}C_{\mu\lambda\nu\kappa},
\end{equation}
is the Bach tensor. A key characteristic of the theory is that~\eqref{eq:BFE} contains fourth-order derivatives, as is evident from the Weyl tensor being covariantly differentiated twice in~\eqref{eq:Bach Tensor}. Meanwhile, GR's Einstein field equations 
\begin{equation}\label{eq:EFE}
    G_{\mu\nu} =8 \uppi \mathrm{G} \, T_{\mu\nu}
\end{equation}
are only second-order, given that the Einstein tensor is $G_{\mu \nu} = R_{\mu \nu}-\frac{1}{2}R \, g_{\mu \nu}$. Here, ${\rm G}$ is the Newtonian gravitational constant, and we take the speed of light to be $\mathrm{c}=1$. 

It should be noted that while $T_{\mu \nu}$ appears in both~\eqref{eq:BFE} and~\eqref{eq:EFE}, the profiles of the $T_{\mu \nu}$ components will not necessarily be identical for the same given mass-energy distribution in the action. For instance, conformal invariance demands that $T_{\mu \nu}$ be traceless in CG, while this restriction is not found in GR~\cite{makingthecase}. 

While CG and GR are both covariant metric theories of gravity and thus share much of the same formalism, the additional conformal symmetry encoded in the theory and the fourth-order derivative structure of the Bach field equations~\eqref{eq:BFE} necessitate some fundamental differences. Aside from the aforementioned tracelessness of $T_{\mu \nu}$ in CG, the higher-order derivative nature of $W_{\mu \nu}$ means that extra integrations are needed to solve the CG field equations~\eqref{eq:BFE} compared to those of GR~\eqref{eq:EFE}.

In recent years, purported solutions to CG have been proposed by~\cite{TFT, PF, fathi3, fathi2021} via a supposed \textit{weak-field} method. This line of papers begins with the methodology introduced in the work of Tanhayi, Fathi, and Takook (TFT)~\cite{TFT}. With a spherically-symmetric setup with a metric of the form
\begin{equation}\label{eq:sphsymm}
    g_\munu = \text{diag}(g_{00}, g_{rr}, g_{\theta \theta}, g_{\phi \phi})=\text{diag}\left(-B(r), \dfrac{1}{B(r)}, r^2, r^2\sin^2\theta\right),
\end{equation}
they consider the correction $h_{\mu \nu}$ to the Minkowski background metric $\eta_{\mu \nu}$ to be small, such that
 \begin{equation}\label{eq:weak field limit}
     g_{\mu \nu} = \eta_{\mu \nu} + h_{\mu \nu}, \ \ |h_{\mu \nu}| \ll 1.
 \end{equation}
In this weak-field setup, one may decompose the lapse function $B(r)$ as
 \begin{equation}\label{eq:LapseDecomposition}
     B(r) = 1 + 2 \, \Phi(r),
 \end{equation}
 where $\Phi(r)$ is the gravitational potential.
 
 They claim to produce a weak-field solution to CG~\eqref{eq:BFE} for a homogeneous sphere with a second-order Poisson equation obtained from the Einstein equations~\eqref{eq:EFE}:
 \begin{equation}\label{eq:2nd Order Poisson}
     \nabla^2 B(r) = 8 \uppi\mathrm{G} \, T_{00}. 
 \end{equation}
From~\eqref{eq:LapseDecomposition}, this is equivalent to $\nabla^2 \Phi(r) = 4 \uppi \mathrm{G} \rho$, with $\rho$ being the density. This is where the main conceptual problem lies. It is evident that this is a weak-field version of Einstein's second-order derivative field equations~\eqref{eq:EFE} and not of CG's fourth-order derivative field equations~\eqref{eq:BFE}. Furthermore, a crucial sign error is made in the method of~\cite{TFT}, and they end up actually solving $ \nabla^2( -B(r)) = 8 \uppi\mathrm{G} \, T_{00}. $
 
 While it may be tempting to assume that reduction of the field equations~\eqref{eq:BFE} to a second-order Poisson equation is one of the aspects of GR carried over to CG, the manifestly fourth-order derivative structure of the Bach tensor $W_{\mu \nu}$~\eqref{eq:Bach Tensor} prevents this from being the case. In spherical symmetry, the Bach equations instead give rise to a fourth-order Poisson equation~\cite{mannheimgammavalue, makingthecase} of the form
\begin{equation}\label{eq:4th_poisson}
    \nabla^4B(r)=\dfrac{3({T^0}_0-{T^r}_r)}{4\alpha_\mathrm{g} B(r)},
\end{equation}
which is obtained by evaluating ${W^0}_0-{W^r}_r=({T^0}_0-{T^r}_r)/4\alpha_\mathrm{g}$ from~\eqref{eq:BFE} and rearranging appropriately. 

Using the weak-field parametrisation of~\eqref{eq:LapseDecomposition}, one can reduce~\eqref{eq:4th_poisson} to $\nabla^4 \Phi(r)=h(r) =-3(\rho/B+P_r\,B)/(8\alpha_\mathrm{g}B)$ with $h(r)$ being the source and $P_r$ the radial pressure~\cite{mannheimgammavalue}. This in turn goes to $h(r)\sim\rho+P_r$ in the weak-field limit where $B\approx1$, as $\alpha_\mathrm{g}<0$~\cite{mannheim_2007,KeithRotation}. Here, the radial pressure is $T_{rr}=P_r$ such that ${T^r}_r=T_{rr}g^{rr}={P_r}\,B$. When pressure can be neglected, $\nabla^4 \Phi(r) \propto \rho$.

In this work, we demonstrate that the method of generating weak-field solutions to CG using a second-order Poisson equation~\cite{TFT} is mistaken. Thus, the proposed metrics using this method representing an uncharged homogeneous sphere~\cite{TFT}, charged homogeneous sphere~\cite{PF, fathi3}, and charged rotating source~\cite{fathi2021} are invalid solutions to CG. Our goal here is not to generate correct weak-field solutions, but to show that the ones mentioned are erroneous. Some valid weak-field solutions to CG may be found in~\cite{mannheimgammavalue, Brihaye_2009}. Furthermore, by stressing the importance of always considering the full fourth-order nature of CG, we hope to prevent the continual proliferation of works using second-order techniques for this theory.

This fundamental conceptual error is compounded by the application of known (electro)vacuum solutions to CG to non-vacuum setups. We therefore begin in~\cref{sec:Solutions} by reviewing the known (electro)vacuum solutions to CG in order to better situate the TFT method and solutions generated from it. We proceed to describing the TFT method and the uncharged homogeneous sphere in~\cref{sec:TFT}. After this, we deal with charged solutions generated with the TFT method in~\cref{sec:Charged}. Finally, we conclude in~\cref{sec:Conclusions}. 

\section{(Electro)vacuum solutions to Conformal Gravity}\label{sec:Solutions}

The static spherically symmetric vacuum solution to GR~\eqref{eq:EFE} is given by the GR Schwarzschild (GRS) metric, which has a lapse function 
\begin{equation}\label{eq:GRS}
   B_{\text{GRS}}(r) = 1- \dfrac{2\mathrm{G}M}{r}.
\end{equation}

When adding in an electromagnetic vector potential, valid exterior to a charge $Q$, of the form
\begin{equation}\label{eq: vector potential}
    A_\mu = \left( \dfrac{Q}{r},0,0,0\right),
\end{equation}
one gets the electrovacuum static, spherically symmetric solution known as the GR Reissner-Nordstr\"{o}m (GRRN) metric given by
\begin{equation}\label{eq:GRRN} 
    B_{\mathrm{GRRN}}(r)=1-\dfrac{2\,\mathrm{G}M}{r}+\frac{\mathrm{G}Q^2}{r^2},
\end{equation} 
for a stress-energy tensor unnormalised by $1/4\uppi$. 

Meanwhile, exact solutions in CG for the (electro)vacuum static, spherically symmetric case were found by Riegert in 1984~\cite{Riegert1984}. These have lapse functions of the form
\begin{equation}\label{eq:sphsymm_Ansatz}
    B(r)=w+\dfrac{u}{r}+vr-kr^2,
\end{equation}
and $w$, $u$, $v$, $k$ are integration constants. This can be obtained as a solution to the homogeneous fourth-order Poisson equation with $\nabla^4B(r)=0$~\eqref{eq:4th_poisson}. To ensure that this parametrization simultaneously solves both the $00$ and $rr$ components of Bach field equations~\eqref{eq:BFE}, one may check that the coefficients satisfy the third-order constraint from the $rr$ equation given by
\begin{equation}\label{eq:3_order}
    \dfrac{1+3uv-w^2}{3r^4}=\dfrac{{T^r}_r}{4\alpha_\mathrm{g}}.
\end{equation}
This constraint holds for both the uncharged and charged cases, and thus
\eqref{eq:sphsymm_Ansatz} represents the CG analogue
to the GR Schwarzschild and Reissner-Nordstr\"{o}m metrics. 

In deriving this solution, Riegert also provided a CG analogue to Birkhoff's Theorem~\cite{JebsenBirkhoff, Birkhoff1923}, thus showing that this is the unique (electro)vacuum spherically symmetric solution in CG~\cite{Riegert1984}. Crucially, the coefficients $w, u, v, $ and $k$ are only necessarily constants in (electro)vacuum. When applied to non-(electro)vacuum solutions, such as in the interior of an energy distribution, these coefficients may become functions of $r$~\cite{Brihaye_2009, kusano2026_shells}.

In 1989, Mannheim and Kazanas restated~\eqref{eq:sphsymm_Ansatz} for the CG Schwarzschild analogue~\cite{mannheim1989}  as
\begin{equation}\label{eq:MK_metric}
    B_{\mathrm{MKS}}(r)=(1-3\,\beta\,\gamma) -\dfrac{\beta\,(2-3\,\beta\,\gamma)}{r}+\gamma\, r-\kappa\, r^2 ,
\end{equation}
with integration constants $\beta$, $\gamma$, and $\kappa$.
In a vacuum where ${T^r}_r=0$, it is clear that the parametrisation of~\eqref{eq:MK_metric} satisfies the third-order constraint~\eqref{eq:3_order}. We shall refer to~\eqref{eq:MK_metric} as the MK-Schwarzschild (MKS) metric. It is this form of the metric that has been used to fit the parameters $\beta$, $\gamma$ and $\kappa$ to rotation curves and cosmological curvature~\cite{mannheim1989, mannheim2011, Mannheim2012Rot, Mannheim2013Rot, KeithRotation,VarieschiParameters1, mannheimgammavalue, VarieschiCosmo, mannheim_2022}. 

In the limit $\gamma, \kappa \rightarrow 0$, one recovers the GRS solution, with $\beta$ serving as the mass-like parameter in this GR limit. 
Since $\gamma$ and $\kappa$ are relatively 
small, solar system phenomenology is mostly unaffected by the CG corrections~\cite{mannheimgammavalue, makingthecase,mannheim_2007}. 

Mannheim and Kazanas later wrote the charged Reissner-Nordstr\"{o}m solution to CG~\cite{mannheim1991} given a vector potential of the form~\eqref{eq: vector potential}.
This MKRN metric has a lapse function given by
\begin{equation}\label{eq:CGRN_metric_1}
    B^{\gamma \neq0}_{\mathrm{MKRN}}(r)=\left( 1-3\,\beta\,\gamma \right)
    -\dfrac{1}{r}\,\left(\beta\,(2-3\,\beta\,\gamma) +\dfrac{Q^2}{8\, \alpha_\mathrm{g}\,\gamma}\right)+\gamma \,r-\kappa\, r^2, 
\end{equation}
when $\gamma \neq 0$ \footnote{ The metric for the case of $\gamma = 0$ has a different form, which is needed due to the degeneracy in~\eqref{eq:CGRN_metric_1} when $\gamma = 0$. However, as this is a special case, we shall focus on the $\gamma \neq 0$ solution~\eqref{eq:CGRN_metric_1} and instead point readers to the discussions in~\cite{mannheim1991, kusano2025_CGRN}. }. 

It is clear that the MKS~\eqref{eq:MK_metric} and MKRN~\eqref{eq:CGRN_metric_1} solutions are of the form of~\eqref{eq:sphsymm_Ansatz}. 
 Interestingly, while the MKS metric~\eqref{eq:MK_metric} reduces to GRS when $\gamma, \kappa \rightarrow 0$, the MKRN~\eqref{eq:CGRN_metric_1} metric does not reduce to the GRRN solution~\eqref{eq:GRRN}. This is evident from the fact that the charge $Q$ is found in the $1/r$ term in the CG solution~\eqref{eq:CGRN_metric_1}, while it is found in the $1/r^2$ term in GR. The presence of the linear $\gamma r$ and quadratic $\kappa r^2$ terms in~\eqref{eq:MK_metric} and~\eqref{eq:CGRN_metric_1} highlights the fourth-order derivative nature of CG as opposed to the second-order nature of GR.

We can now discuss the additional problem that arises in the TFT method. Aside from erroneously using a second-order Poisson equation~\eqref{eq:2nd Order Poisson} instead of the appropriate fourth-order one~\eqref{eq:4th_poisson} in CG, they assume that solutions to non-(electro)vacuum setups, where $\rho \neq 0$, are also described by~\eqref{eq:sphsymm_Ansatz} with constant coefficients. However, as we have mentioned, in the interior of a source, the coefficients $w, u, v,$ and $k$ may be functions of $r$. Thus, we show that they presuppose the form of the metric and apply it to a setup where it is not valid.

\section{TFT weak-field method and the uncharged homogeneous sphere}\label{sec:TFT}

We begin this section with an overview of the weak-field method developed by TFT by going through their derivation of their metric representing an uncharged homogeneous sphere, which we shall refer to as the TFT metric. We then move on to dealing with the issues in their method in the succeeding subsections.

\subsection{Derivation of the TFT metric and summary of the TFT method}

In~\cite{TFT}, they consider a setup with a point mass $M$ immersed in a homogeneous sphere of density $\rho_0 = 3M_0/4 \uppi R_0^3$. Given their weak-field limit~\eqref{eq:weak field limit}, they write $g_{00}=\eta_{00}+ h_{00}=-B(r)$. 

They take an ansatz for a metric of the form 
\begin{equation}\label{eq:metric ansatz}
    B_\mathrm{TFT}(r) = 1-\dfrac{2 \mathrm{G}M}{r}-\dfrac{1}{3}f(r)r^2,
\end{equation}
for a function $f(r)$ such that
\begin{equation}\label{eq:tft_ansatz}
    f(r) = -c_1 + \dfrac{6 \mathrm{G}M}{r^3},
\end{equation}
with an arbitrary constant $c_1$. Thus, the lapse function mimics the Schwarzschild-(Anti)-de Sitter ((A)dS) metric of GR as
\begin{equation}\label{eq: TFT Metric}
   B_{\text{TFT}}(r) 
   =1- \dfrac{4\mathrm{G}M}{r} +\dfrac{1}{3}c_1r^2.
\end{equation}
This is supposedly motivated by the fact that a metric of the form of Schwarzschild-(A)dS is a vacuum solution to CG~\cite{mannheim1989}\footnote{There is a known correspondence between CG and GR in the specific case of GR-(A)dS spacetimes, where the application of Neumann boundary conditions to impose spontaneous symmetry breaking can make CG equivalent to second-order GR~\cite{Maldacena:2011mk, Anastasiou_2016, Anastasiou_2021}. However, the cases being considered here are evidently non-vacuum cases, so such an equivalence does not hold.}. 
From this, one would have de Sitter (dS) backgrounds when $c_1 < 0$, and Anti-de Sitter (AdS) backgrounds when $c_1>0$.

They~\cite{TFT} then move on to taking a second-order Poisson equation as
\begin{equation}\label{eq:TFT Second Order Neg}
\left(\dfrac{\mathrm{d}^2}{\mathrm{d} r^2}+\dfrac{2}{r} \dfrac{\mathrm{d}}{\mathrm{d} r}\right)\left(-1+\dfrac{4 \mathrm{G} M}{r}- \dfrac{1}{3}c_1 r^2\right)= 8 \uppi \mathrm{G} T_{00} = 8 \uppi \mathrm{G} \rho_0
.
\end{equation}
Looking at~\eqref{eq:LapseDecomposition} and~\eqref{eq: TFT Metric}, we see that this in fact has the wrong sign for a second-order Poisson equation~\eqref{eq:2nd Order Poisson}. They solve for the arbitrary constant to get
\begin{equation}\label{eq:TFT c1 wrong}
    c_1 = -4 \uppi \mathrm{G} \, \rho_0  = -3\dfrac{\mathrm{G}M_0}{R_0^3}, 
\end{equation}
and
\begin{equation}\label{eq: TFT Solved wrong}
   B_{\text{TFT}}(r) =1- \dfrac{4\mathrm{G}M}{r} -\dfrac{\mathrm{G}M_0}{R_0^3}r^2.
\end{equation}
Due to the sign error, they erroneously find a dS background with $c_1 < 0$, and so take $c_1$ as an estimate for the cosmological constant $\Lambda$, by considering $M_0$ and $R_0$ to be the mass and radius of the observable universe. 

If one instead takes the correct sign for the second-order Poisson equation~\eqref{eq:2nd Order Poisson}, 
\begin{equation}\label{eq:TFT Second Order}
\left(\dfrac{\mathrm{d}^2}{\mathrm{d} r^2}+\dfrac{2}{r} \dfrac{\mathrm{d}}{\mathrm{d} r}\right)\left(1-\dfrac{4 \mathrm{G} M}{r}+ \dfrac{1}{3}c_1 r^2\right)= 8 \uppi \mathrm{G} T_{00} = 8 \uppi \mathrm{G} \rho_0
,
\end{equation}
one would get
\begin{equation}\label{eq: TFT Solved}
   B_{\text{TFT}}(r) =1- \dfrac{4\mathrm{G}M}{r} +\dfrac{\mathrm{G}M_0}{R_0^3}r^2.
\end{equation}
Thus, with $c_1 > 0$, this gives an Anti-de Sitter (AdS) background. 

We can here summarize the steps involved in TFT weak-field method~\cite{TFT}.

\begin{enumerate}
    \item Take an ansatz for $f(r)$ and thus the lapse function $B(r)$ that takes the form of the (electro)vacuum CG solution~\eqref{eq:sphsymm_Ansatz} for some arbitrary coefficients $c_i$.

    \item Assume a weak-field limit for the metric $g_{\mu \nu}$~\eqref{eq:weak field limit}.

     \item Apply their $B(r)$, valid for (electro)vacuum where $\rho = 0$, to a setup in the interior of a matter distribution where $\rho \neq 0$.

    \item Use a second-order Poisson equation~\eqref{eq:TFT Second Order Neg} (of the wrong sign) on the lapse function $B(r)$ where the RHS is from the GR energy-momentum tensor $T_{\mu \nu}$.

    \item Solve for the coefficients $c_i$, assuming that they are constants.
\end{enumerate}

The primary conceptual problem in this method lies in Step (iv), as second-order instead of fourth-order derivatives of the metric are being used to solve for a solution to CG~\eqref{eq:BFE}. It is furthermore clear that \eqref{eq:TFT Second Order Neg} is the RHS of the Einstein Field Equations of GR \eqref{eq:EFE}, and not the appropriate one from CG. We tackle this in Subsection \ref{subsection: fourth order}.

The next main issue, seen in Step (iii), is the use of a valid (electro)vacuum $(\rho = 0)$ ansatz for $B(r)$ in \eqref{eq:metric ansatz} to \eqref{eq: TFT Metric} in a setup that is not source-free $(\rho \neq0)$. This leads to inconsistencies in the form of the stress-energy tensor $T_{\mu \nu}$ in Step (iv), and this cascades into a spurious evaluation of the coefficients of $B(r)$. This is discussed in Subsection \ref{subsection: electrovacuum}.

\subsection{Need for a fourth-order Poisson equation}\label{subsection: fourth order}

One may be inclined to ask if the use of the second-order Poisson equation~\eqref{eq:2nd Order Poisson} may be valid in the weak-field limit~\eqref{eq:weak field limit} of CG, even if it does not hold in the full theory. However, when one linearizes the Bach tensor~\eqref{eq:Bach Tensor} about the Minkowski background $\eta_{\mu \nu}$~\eqref{eq:weak field limit}, one gets
\begin{equation}
    \delta W_{\mu \nu} = 2\, \delta\left(\nabla^\kappa\nabla^\lambda C_{\mu\lambda\nu\kappa}\right)
-
\delta R^{\lambda\kappa}\overline{C}_{\mu\lambda\nu\kappa} - \overline{R}^{\lambda\kappa} \delta C_{\mu\lambda\nu\kappa} = 2\, \delta\left(\nabla^\kappa\nabla^\lambda C_{\mu\lambda\nu\kappa}\right),
\end{equation}
with the background quantities
$\overline{C}_{\mu\lambda\nu\kappa}   =\overline{R}^{\lambda\kappa}=0$ vanishing in Minkowski. Clearly, the second-order derivatives of $C_{\mu\lambda\nu\kappa}$ remain, and thus we still have fourth-order derivatives of $h_{\mu \nu}$ overall, as similarly found by~\cite{RiegertLinearized}.

This then conclusively demonstrates that a proper solution to CG in the weak-field limit requires a fourth-order Poisson equation, not a second-order one. Instead,~\eqref{eq:4th_poisson} would yield something of the form
\begin{equation}\label{eq:Spherical 4th}
\left(\dfrac{\mathrm{d}^2}{\mathrm{d} r^2}+\dfrac{2}{r} \dfrac{\mathrm{d}}{\mathrm{d} r}\right)\left(\dfrac{\mathrm{d}^2}{\mathrm{d} r^2}+\dfrac{2}{r} \dfrac{\mathrm{d}}{\mathrm{d} r}\right)B(r)\propto  \rho_0.
\end{equation}
We can tell from this that the TFT method is missing two further integrations in determining the coefficients of solutions. 

The reason why the CG MKS metric~\eqref{eq:MK_metric} works for solar system tests~\cite{mannheim_2007} is not because CG becomes second-order in the weak-field limit. Rather, as mentioned earlier, the $\gamma$ and $\kappa$ parameters are sufficiently small that the linear $\gamma r$ and quadratic $-\kappa r^2$ potentials do not significantly contribute at solar system distance scales, and one thus reproduces results of the GRS metric~\eqref{eq:GRS}.

\subsection{Invalid use of the (electro)vacuum ansatz}\label{subsection: electrovacuum}

While a metric of the form of~\eqref{eq: TFT Metric} does technically solve the vacuum Bach equations~\eqref{eq:BFE} and does satisfy the vacuum third-order constraint~\eqref{eq:3_order} where $T^r_r = 0$, the setup being considered here is not vacuum. Instead, it is the interior of a homogeneous sphere of density $\rho_0$. As mentioned earlier, in non-(electro)vacuum setups, the coefficients of~\eqref{eq:sphsymm_Ansatz} are not necessarily constants, and may in fact become functions of $r$ as $w(r), u(r), v(r)$ and $k(r)$~\cite{Brihaye_2009, kusano2026_shells}. These are obtained as moment integrals of the source $\nabla^4 \Phi(r)=h(r)$~\eqref{eq:4th_poisson}, such that~\cite{mannheim_2007, KeithRotation, kusano2026_shells}: 
\begin{equation}\label{eq:moments_w_boundary} 
    \begin{aligned} 
        w'(r) = +r^3 h(r),\hspace{0.4cm}
        u'(r) = \frac{-r^4}{3} h(r),\hspace{0.4cm}
        v'(r) = -r^2 h(r),\hspace{0.4cm}
        k'(r) = \frac{-r}{3} h(r),
    \end{aligned}
\end{equation} 
with primes indicating derivatives of $r$.  Boundary conditions of $u$ and $v$ are set at $r=0$, while the boundary condition of the cosmological $k$ is set at $r\rightarrow+\infty$. One can then choose the boundary condition for $w$ to be set at either $r=0$ or $r\rightarrow+\infty$, provided that~\eqref{eq:3_order} is satisfied~\cite{Brihaye_2009}. 

Matter-sourced solutions of CG are then generated through integrating~\eqref{eq:moments_w_boundary} alongside the appropriate equations of state associated with the matter fields. We remark that crucially, this also applies to weak-field solutions with extended matter distributions.

One must thus be careful not to presuppose the form of the metric and its coefficients when considering a non-(electro)vacuum setup in CG. As $c_1$ could be a function of $r$, the evaluation of the Poisson equation may have required the radial derivatives to act on $c_1(r)$. This issue becomes more apparent in the next section, where we deal with the charged homogeneous sphere of Payandeh and Fathi~\cite{PF}.

We will recall that TFT place a point mass $M$ at the origin of the homogeneous sphere. Another inconsistency is then apparent when we set the density of the homogeneous sphere to vanish $\rho_0=0$ in the vacuum limit. The TFT metric~\eqref{eq: TFT Solved wrong} yields
\begin{equation}
   B_{\text{TFT}, \rho_0=0}(r) = 1- \dfrac{4\mathrm{G}M}{r}.
\end{equation}
As we can see, this does not reproduce the GRS solution~\eqref{eq:GRS}, as the $1/r$ term appears to be too high by a factor of $2$. A reparametrization of $M = 1/2 \, M_{\text{physical}}$ would fix this. This reparametrization can be done given that in this vacuum limit, the third-order constraint~\eqref{eq:3_order} does not specify the coefficient of the $1/r$ term. One may then turn to such physical considerations to set it properly.

%\rlk{The TFT metric has been interpreted as considering density contribution from a cosmological constant added into the Bach equations \cite{TFT, PF}. This would subsequently be in good agreement with their physical interpretation for $M_0$ and $R_0$, and aligns well with why these two parameters appear in the (Anti-)de Sitter term of the lapse. } 

%\rlk{Nevertheless, such an interpretation would still be problematic, even if one were to ignore the fact that the wrong field equations are integrated. For one, a cosmological constant added into the field equations via $\Lambda g_\munu$ term or similar, without the presence of other counteracting terms, would violate the tracelessness condition that permits the conformal invariance of CG in the first place. Furthermore, a significant appeal of the present theory is that a cosmological constant is simply not required in the first place to get a quadratic (A)dS term, but arises directly as a result of integrating the higher-order field equations. } 

\section{Charged solutions}\label{sec:Charged}
In this section, we discuss the Payandeh and Fathi (PF)~\cite{PF} metric for a charged homogeneous sphere and the charged rotating metric of Fathi, Olivares, and Villanueva (FOV)~\cite{fathi2021}. Being generated from the TFT method \cite{TFT}, we point out how they similarly end up erroneous.

\subsection{Payandeh and Fathi's charged homogeneous sphere}
Using the weak-field TFT method~\cite{TFT}, PF~\cite{PF} sought to find a solution for a charged homogeneous sphere of density $\rho_0 = 3M_0/4 \uppi R_0^3$ and total charge $Q$. 

Following Step (i) of the TFT method~\cite{TFT}, they take a function
\begin{equation}\label{eq:PF f(r)}
    f(r) = -c_1  - \dfrac{c_2}{r} - \dfrac{6 \mathrm{G}M}{r^3},
\end{equation}
to give a lapse function of the form
\begin{equation}\label{eq: PF Metric}
   B_{\text{PF}}(r) = 1-\dfrac{2 \mathrm{G}M}{r}-\dfrac{1}{3}f(r)r^2= 1+\dfrac{1}{3}c_2r +\dfrac{1}{3}c_1r^2,
\end{equation}
where, as in Step (ii), they assume that this is a weak-field metric~\eqref{eq:weak field limit}. We may observe, upon comparing to the general form of the TFT metric~\eqref{eq: TFT Metric}, that by their choice of $f(r)$ they have done away with the central mass $M$ in the lapse function $B_{\text{PF}}(r)$. They now also have an additional linear term $c_2 r$. It is unclear why this difference in choice from the TFT metric~\cite{TFT} is made.

As with the previous case, they take the assumption of Step (iii) that~\eqref{eq: PF Metric} takes the form of (electro)vacuum metrics in CG, and so assume that the lapse function still takes on constant coefficients in this non-(electro)vacuum setup where $\rho \neq 0$.

Now, for Step (iv), the sign error in the TFT method~\cite{TFT} carries over and they take
\begin{equation}\label{eq:PF Second Order}
\left(\dfrac{\mathrm{d}^2}{\mathrm{d} r^2}+\dfrac{2}{r} \dfrac{\mathrm{d}}{\mathrm{d} r}\right)\left(-1-\dfrac{1}{3}c_2r -\dfrac{1}{3}c_1r^2\right)= 8 \uppi \mathrm{G}\left( T_{00}+E_{00}\right),
\end{equation}
where $T_{00} = \rho_0$ as earlier, and $E_{00}$ is supposedly the stress-energy tensor component derived from a vector potential $A_\mu$ as given in~\eqref{eq: vector potential}. They give this to be

\begin{equation}
    E_{00}=\dfrac{Q^2}{8\uppi\,r^4}.
\end{equation}

Aside from the aforementioned sign error, a few issues immediately arise here. First,  ~\eqref{eq:PF Second Order} wrongly assumes that the coupling constant between $T_\munu$ and the curvature tensor in the field equations is also $8\pi {\rm G}$ in CG as it is in GR. This is immediately evident from the Bach field equations~\eqref{eq:BFE}, where $T_{\mu \nu}$ is multiplied by the dimensionless coupling constant $\alpha_\mathrm{g}$. This is why in the MKRN metric~\eqref{eq:CGRN_metric_1}, the charge term has a factor of $\alpha_\mathrm{g}$. Meanwhile, in~\eqref{eq:PF Second Order}, $E_{00}$ is multiplied by $\mathrm{G}$ as in GR~\eqref{eq:EFE}. 

Secondly, the $Q/r$ behavior in the vector potential $A_\mu$~\eqref{eq: vector potential} is valid exterior to the total charge $Q$. However, as may be seen in~\eqref{eq: PF Metric}, they are taking $T_{00}=\rho_0\neq 0$. This means they are supposedly evaluating this second-order Poisson equation within the homogeneous charged sphere. This is inconsistent.

Evaluating~\eqref{eq:PF Second Order}, they arrive at
\begin{equation}\label{eq: PF c1}
c_1=
-\dfrac{1}{2}\dfrac{\mathrm{G}Q^2}{r^4}
-\dfrac{1}{3}\dfrac{c_2}{r}-\dfrac{3\mathrm{G}M_0}{R_0^3},
\end{equation}
and
\begin{equation}\label{eq: PF c2}
c_2=
-\dfrac{3}{2}\dfrac{\mathrm{G}Q^2}{r^3}-\dfrac{9\mathrm{G}M_0}{R_0^3}r
-3c_1 r.
\end{equation}
The problem with Steps (iii) and (v) of the TFT method is immediately seen here. Contrary to their assumption of constant coefficients in their ansatz for $f(r)$~\eqref{eq:PF f(r)} and $B(r)$~\eqref{eq: PF Metric}, $c_1$ and $c_2$ are not constants, but are instead functions of $r$. They should thus be written as $c_1=c_1(r)$ and $c_2=c_2(r)$. This would have changed the evaluation of~\eqref{eq:PF Second Order}, with the derivatives with respect to $r$ acting on $c_1$ and $c_2$. It is also questionable that these coefficients were not subjected to boundary conditions in determining them.

However, this inconsistency is not recognized by PF~\cite{PF}, and they instead write out two different lapse functions. The first uses $c_1$ as given in~\eqref{eq: PF c1} to give
\begin{equation}\label{eq: PF metric 1 wrong}
    B_{\text{PF,1}}(r)=1
-\dfrac{1}{6}\dfrac{\mathrm{G}Q^2}{r^2}
+\dfrac{2}{9}c_2 r-\dfrac{\mathrm{G}M_0}{R_0^3}r^2.
\end{equation}
The second uses $c_2$~\eqref{eq: PF c2} to get
\begin{equation}\label{eq: PF metric 2 wrong}
    B_{\text{PF,2}}(r)=1
-\dfrac{1}{2}\dfrac{\mathrm{G}Q^2}{r^2}+\left(
-\dfrac{2}{3}c_1 -\dfrac{3\mathrm{G}M_0}{R_0^3}\right)r^2.
\end{equation}

If one uses a second-order Poisson equation of the correct sign~\eqref{eq:TFT Second Order}, but still disregards that $c_1$ and $c_2$ are functions of $r$, one would instead arrive at 
\begin{equation}\label{eq: PF metric 1 correct}
    B_{\text{PF,1}}(r)=1
+\dfrac{1}{6}\dfrac{\mathrm{G}Q^2}{r^2}
+\dfrac{2}{9}c_2 r+\dfrac{\mathrm{G}M_0}{R_0^3}r^2,
\end{equation}
and
\begin{equation}\label{eq: PF metric 2 correct}
    B_{\text{PF,2}}(r)=1+
\dfrac{1}{2}\dfrac{\mathrm{G}Q^2}{r^2}
+\left(
-\dfrac{2}{3}c_1 +\dfrac{3\mathrm{G}M_0}{R_0^3}\right)r^2.
\end{equation}
We thus notice the discrepancy in the signs of terms in these lapse functions. 

One observes that the charge terms in~\eqref{eq: PF metric 1 wrong} and~\eqref{eq: PF metric 2 wrong}  go as $1/r^2$ as in the GRRN~\eqref{eq:GRRN} lapse function. Importantly, however, due to the aforementioned sign error they differ in the sign. While the gravitational effect of the charge term in GR is repulsive, it is here attractive in the PF metrics.

We may also notice that they differ from the $1/r$ behavior we see in the MKRN metric in CG~\eqref{eq:CGRN_metric_1}. In fact, there is no $1/r^2$ term at all in~\eqref{eq:sphsymm_Ansatz}, the lapse function for exact electrovacuum external solutions of CG. Of course, we have already pointed out that the physical setup of the PF metric~\cite{PF} is unclear. The $\sim {\mathrm{G}M_0r^2}/{R_0^3}$ term in~\eqref{eq: PF metric 1 wrong} and~\eqref{eq: PF metric 2 wrong} hints that this metric is an internal solution of a charged homogeneous sphere. However, the vector potential $A_\mu$~\eqref{eq: vector potential} used is meant for evaluation outside a charged source.

Regardless of these individual inconsistencies, the fact that a second-order Poisson equation is used at all as a weak-field limit solution to the fourth-order Bach equations~\eqref{eq:BFE} ensures that the PF solution~\cite{PF} is not a valid CG solution. One can note that while ${G^\mu}_\nu$ and ${W^\mu}_\nu$ have dimensions of $1/r^{2}$ and $1/r^4$ respectively, the components of ${T^\mu}_\nu$ for electrovacuum metrics in both theories are proportional to $1/r^4$; this leads to the charge term following different radial profiles in CG and GR. The uniqueness theorem of Riegert~\cite{Riegert1984} also ensures that no $Q^2/r^2$ term is seen in the MKRN solution. 

It is also rather strange that this solution, generated via a supposed weak-field limit~\eqref{eq:weak field limit}, has been analyzed as a black hole solution~\cite{PF}. For instance,~\cite{fathi3, fathi2020, fathi2} even analyze the horizon structure of such black holes, where the limit $|h_{\mu \nu}|\ll 1$ would no longer hold.

For future discussions on nonrotating charged black holes in CG, we recommend readers to instead use~\eqref{eq:CGRN_metric_1}. For a more extensive discussion of the MKRN metric, we refer readers to~\cite{kusano2025_CGRN}. 

\subsection{Fathi, Olivares, and Villanueva's charged rotating source}

FOV~\cite{fathi2021} claim to generate a charged rotating solution to CG using a modified Newman-Janis algorithm~\cite{NewmanJanis,ModifiedNewmanJanis1, ModifiedNewmanJanis2}. This is a method of arriving at a rotating solution by a complex coordinate transformation of a known correct static solution.

Beginning with the charged homogeneous sphere of the PF metric~\cite{PF} in~\eqref{eq: PF metric 2 wrong}, they apply the modified Newman-Janis algorithm~\cite{ModifiedNewmanJanis1, ModifiedNewmanJanis2} to obtain the metric~\cite{fathi2021} written as 
\begin{equation}
\begin{aligned}
{\rm d}s^2 ={}& -\dfrac{\Xi}{\Sigma}\,{\rm d}t^2
+\dfrac{\Sigma}{\Delta}\,{\rm d}r^2
+\Sigma\,{\rm d}\theta^2
-2a\sin^2\theta\left(1-\dfrac{\Xi}{\Sigma}\right){\rm d}t\,{\rm d}\phi \\
&+\sin^2\theta
\left[
\Sigma+a^2\sin^2\theta
\left(
2-\dfrac{\Xi}{\Sigma}
\right)
\right]{\rm d}\phi^2 ,
\end{aligned}
\end{equation}
with auxiliary functions
\begin{equation}
\Delta=a^2+r^2B_{\text{PF,2}}(r),\qquad
\Xi=\Delta-a^2\sin^2\theta,\qquad
\Sigma=r^2+a^2\cos^2\theta.
\end{equation}

It is worth noting that~\cite{fathi2021} does not state whether the modification applied to the Newman-Janis algorithm is compatible with the present CG theory. Nevertheless, given that the basis static metric used for the Newman-Janis algorithm is the erroneous PF metric~\eqref{eq: PF metric 2 wrong} from~\cite{PF} as discussed in the previous subsection, it is easy to write off this rotating counterpart of FOV~\cite{fathi2021} as likewise being in error.

Mannheim and Kazanas~\cite{mannheim1991} derived the exact uncharged rotating and charged rotating (electro)vacuum solutions to the Bach field equations~\eqref{eq:BFE} of CG. While more comprehensive studies of the charged rotating case have yet to be done, analyses of the uncharged rotating solution may be found in~\cite{VarieschiFlyby, Varieschishadow, yulo2025}.

\section{Conclusions}\label{sec:Conclusions}

In this work, we have discussed the fourth-order nature of Weyl Conformal Gravity and its Bach field equations. We note that in the weak-field limit, it reduces  to a fourth-order Poisson equation. This is unlike the second-order theory of General Relativity, whose Einstein field equations reduce to a second-order Poisson equation. It is the intention of this work to stress this fact in order to help prevent the continual propagation of works using second-order methods for higher-derivative theories such as Conformal Gravity.

By critically analyzing the work of Tanhayi, Fathi, and Takook~\cite{TFT}, we showed that their purported weak-field method of generating Conformal Gravity solutions is invalid, due to it being based on a second-order Poisson equation instead of a fourth-order one. Moreover, we discussed how a valid (electro)vacuum ansatz has been misapplied to non-(electro)vacuum setups.

We further demonstrated that the application of this method to a charged homogeneous sphere in the work of Payandeh and Fathi~\cite{PF} was therefore also erroneous, along with pointing out further inconsistencies in the work. The fact that this metric is invalid also entails that the charged rotating solution of Fathi, Olivares, and Villanueva~\cite{fathi2021} generated from the nonrotating solution of~\cite{PF} via a modified Newman-Janis algorithm is likewise incorrect.

Works based on the weak-field method~\cite{TFT} and corresponding metrics~\cite{TFT, PF, fathi2021} include~\cite{fathi1, fathi2020, fathi2,  QNM, Becar}. These works then all similarly fall in error. 

Additionally, \cite{Jawad:2020wlg, Schmidt_2012, Ghanaatian:2018vzs, candlish2018photonpathshyperbolictopological, Eichhorn_2021, Pantig_2022, Younsi_2023, Meng_2022, patel2022rotationalenergyextractionkerr, Chen:2022dgn, Zubair_2022, Zubair:2023cep, Khan:2023qsl, balali2024deflectionangleshadowslowly, Khan:2024rjr, Fathi:2024sud, Jizba:2025bjq, Momennia:2025ngm, ladino2025phasetransitionsshadowsmicrostructure, Heydari-Fard:2025bdp, Hassanabadi:2026jak, Khan:2026jxt} reference the methods and results of ~\cite{TFT, PF, fathi2021}. While the conclusions of these works do not appear to be crucially dependent on~\cite{TFT, PF, fathi2021}, this highlights the fact that the validity of~\cite{TFT, PF, fathi2021} is continually being taken for granted.

\bmhead{Acknowledgements}

We would like to thank Keith Horne, Jack Purllant, Fredric Hancock, and Giorgos Anastasiou for their insights on this work. We are also grateful for the hospitality of the Centro de Ciencias de Benasque, where some of the discussions regarding this work took place during the workshop ``New Frontiers in Strong Gravity IV.''


\bibliography{sn-bibliography}% common bib file

\end{document}